\documentclass[lettersize,journal]{IEEEtran}
\usepackage{amsmath,amsfonts}
\usepackage{algorithm}
\usepackage{array}
\usepackage[caption=false,font=normalsize,labelfont=sf,textfont=sf]{subfig}
\usepackage{textcomp}
\usepackage{stfloats}
\usepackage{url}
\usepackage{verbatim}
\usepackage{algcompatible}
\usepackage{graphicx}
\usepackage{subcaption}
\usepackage[utf8]{inputenc}

\usepackage{chngcntr}

\usepackage{booktabs,lipsum}

\makeatletter
\newcommand*{\rom}[1]{\expandafter\@slowromancap\romannumeral #1@}
\makeatother

\usepackage{cite}
\usepackage{algpseudocode}
\begin{document}

\title{Consciousness Driven Spike Timing Dependent Plasticity}

\author{\IEEEauthorblockN{Sushant Yadav\IEEEauthorrefmark{1},
Santosh Chaudhary\IEEEauthorrefmark{1},
Rajesh Kumar\IEEEauthorrefmark{2}, }\\
\IEEEauthorblockA{\IEEEauthorrefmark{1} Department of Mathematics, MNIT Jaipur, Jaipur, India, 302017}\\
\IEEEauthorblockA{\IEEEauthorrefmark{2}Department of Electrical Engineering, MNIT Jaipur, Jaipur, India, 302017}
}




\maketitle

\begin{abstract}
Spiking Neural Networks (SNNs), recognized for their biological plausibility and energy efficiency, employ sparse and asynchronous spikes for communication. However, the training of SNNs encounters difficulties coming from non-differentiable activation functions and the movement of spike-based inter-layer data. Spike-Timing Dependent Plasticity (STDP), inspired by neurobiology, plays a crucial role in  SNN's learning, but its still lacks the conscious part of the brain used for learning. Considering the issue, this research work proposes a Consciousness Driven STDP ($CD$-STDP), an improved solution addressing inherent limitations observed in conventional STDP models. $CD$-STDP, designed to infuse the conscious part as coefficients of  long-term potentiation (LTP) and long-term depression (LTD), exhibit a dynamic nature. The model connects LTP and LTD coefficients to current and past state of synaptic activities, respectively, enhancing consciousness and adaptability. This consciousness empowers the model to effectively learn while understanding the input patterns. The conscious coefficient adjustment in response to current and past synaptic activity extends the model's conscious and other cognitive capabilities, offering a refined and efficient approach for real-world applications. Evaluations on MNIST, FashionMNIST and CALTECH datasets showcase $CD$-STDP's remarkable accuracy of 98.6\%, 85.61\% and 99.0\%, respectively, in a single hidden layer SNN. In addition, analysis of conscious elements and consciousness of the proposed model on SNN is performed. 
\end{abstract}

\begin{IEEEkeywords}
Spike Timing Dependent Plasticity, Image Recognition, Consciousness, Spiking Neural Network, Leaky Integrate-and-Fire Neuron.
\end{IEEEkeywords}

\section{Introduction}
The fundamental distinction between machines and minds lies in their core functions as machine computes, while mind comprehend and possess consciousness \cite{edelman2008computing}. Machines process data through algorithms and calculations, functioning within predefined parameters, whereas minds possess the capacity to understand concepts beyond mere computation. Moreover, minds exhibit consciousness, a complex state of awareness and subjective experience that transcends the capabilities of machines \cite{dreyfus1986mind}. Thus, while machines excel in processing information, it is the inherent understanding and consciousness of minds that truly differentiate them. However, Artificial Intelligence (AI) which is considered as modern machine tries to mimic mind \cite{konar2018artificial}. The core part of AI is considered as artificial neural network (ANN) \cite{vishnukumar2017machine}. ANN are mathematical models of biological neural networks. However, the core computational models of ANN have inherent limitation of energy consumption \cite{wozniak2020deep}. To address this issue, SNNs have emerged as an alternative model. SNNs depart from traditional ANNs by mimicking the brain’s spiking neuron behavior, using discrete, event-driven
communication via spikes, similar to biological systems. The crucial capability of SNN is defined by its learning mechanism i.e. STDP \cite{vigneron2020critical}. 

\par
 STDP is a well-established concept in SNNs for learning, which suggests that the timing of neuronal firing plays a crucial role in modifying the strength of synaptic connections between neurons. But it still misses a crucial part of mind i.e. Consciousness. Consciousness, is a multifaceted phenomenon that encompasses subjective awareness, self-reflection, and the integration of information in brain. There are various theories and models of consciousness present in literature. As in \cite{friedman2023current} presents review of Neural Correlates of Consciousness (NCC) in which Neuroscientists aim to identify the neural processes and brain regions associated with conscious experience. Studies using neuroimaging techniques such as fMRI (functional Magnetic Resonance Imaging) and EEG (Electroencephalography) seek to pinpoint the neural correlates of specific conscious states, such as perception, attention, and self-awareness. In \cite{baars2021global} Global Workspace Theory: Proposed by cognitive scientist Bernard Baars, suggests that consciousness arises from the dynamic interactions of distributed neural networks. According to this theory, information becomes conscious when it enters a global workspace where it can be accessed by multiple cognitive systems, facilitating flexible behavior and higher cognitive functions.  In article \cite{gennaro2004higher} gives an overview about Higher-order Theories of Consciousness. These theories propose that consciousness depends on the brain's ability to represent mental states as objects of awareness. According to higher-order theories, consciousness emerges when the brain generates higher-order representations of its own cognitive processes, allowing for introspection and self-awareness. In \cite{tononi2004information} Integrated Information Theory (IIT) developed by neuroscientist Giulio Tononi, proposes that consciousness arises from the integration of information within the brain. According to IIT, a conscious experience corresponds to a specific pattern of information that is both highly integrated and irreducible. This theory provides a mathematical framework for quantifying consciousness and has led to new insights into its neural basis. There are even more theories and ideas to view consciousness like altered states of consciousness \cite{bundzen2002altered}, philosophical perspectives on consciousness \cite{hedman2017consciousness}, clinical studies on consciousness \cite{young2022ethical} etc.. However among all these IIT is considered the most popular as account of the integration of information may be understood as a response to this problem \cite{tononi2008consciousness}. 

 %%%%%%%%%%%%%
 
 While the exact mechanisms underlying consciousness remain elusive, some theories propose that it emerges from the dynamic interactions between neurons, their synaptic connections and modifications. Also in SNN these synaptic connections and modifications are adjusted by STDP. There are various forms of STDP found in literature, as in  \cite{paredes2019unsupervised}, Additive STDP ($C_a$-STDP) relies solely on the parameter $\Delta t $ (i.e. difference in pre and post synaptic spike), rendering it inherently unstable. This instability necessitates additional constraints to effectively bind synaptic weights. Consequently, this approach tends to yield a skewed distribution, with the majority of weights clustering around the upper and lower bounds \cite{sengupta2015sensitivity}. On the other hand, multiplicative STDP ($C_m$-STDP) introduces a bit more complexity by incorporating the $\Delta t $ parameter with an additional emphasis on proportionality concerning weights. Referred to as soft bounds in this scenario, there is a proportional increase in LTD relative to LTP as the synaptic weight grows larger \cite{sjostrom2010spike}. However, despite its heightened adaptability, this rule also leads to a distribution of weights that leans towards the boundaries \cite{bichler2012extraction}.  \cite{burbank2015mirrored} describes $M$-STDP,  centered on the idea that the time window considered for analyzing the correlation between pre and post synaptic spikes should be positioned around the pre synaptic spikes. This introduces the potential for correlating $\Delta t < 0$ with LTP instead of LTD. \cite{tavanaei2016acquisition} introduces $P$-STDP, under this rule, all learning parameters commence with identical values and are allowed to evolve as the quantity of spikes increases. In this model, the amplification parameters for LTP and LTD are maintained at a 4/3 ratio. The key advantage of P-STDP lies in its resilience to heightened mathematical complexity within the neuron model. Reinforcement STDP ($R$-STDP) introduced by \cite{mozafari2018first}, this method is fundamentally based on Pavlov's conditioning approach to learning, a concept that, when combined with Hebb's, has been demonstrated to contribute to the neurobiological approach to learning. The primary advantage of $R$-STDP lies in the network's ability to learn discriminative features instead of repetitive ones. As mentioned in \cite{burbank2012depression} $Rev$-STDP is a complex form of synaptic plasticity occurring in top-down synapses, involving communication in both feed-forward and feed-backward. It place a significant attention on correlation. Presented in \cite{paredes2019unsupervised} $S$-STDP is an evolved form of $C_m$-STDP. It takes normal weight distribution without the need for explicit bounds. As mentioned in \cite{talaei2020pattern,krunglevicius2016modified} $T$-STDP suggests that the idea for LTP and LTD arise from the triplet relationship between two pre-synaptic and one post-synaptic spikes, in either order. This rule is based on the evolution of potentiation with varying frequencies. As mentioned in \cite{garg2022voltage} VDSP learning rule optimizes synaptic conductance by updating only on the postsynaptic neuron's spike, reducing updates by half compared to standard STDP. As all the models found in literature lacks the conscious part, to improve this gap authors proposes a Consciousness Driven Spike Timing Dependent Plasticity (CD-STDP).\par

$CD$-STDP is a basic framework that explores the potential interaction between consciousness and synaptic plasticity. CD-STDP propose that consciousness may influence the timing and pattern of neuronal activity, thereby modulating synaptic plasticity in the brain. In other words, the conscious experience of an individual also helps in shaping the strength and structure of neural connections, potentially influencing cognitive functions. There are several distinctions between $CD$-STDP and other models of STDP mentioned in literature like:
\begin{enumerate}
    \item \textbf{Focus on Neural Activity vs. Conscious Awareness:} Traditional STDP models primarily focus on the timing of neural activity, highlighting how the precise order of firing between neurons leads to synaptic strengthening or weakening. In contrast, CD-STDP shifts the focus to conscious awareness. It proposes that the subjective experience of consciousness influences the patterns of neural activity alongside timing of firing, which in turn modulate synaptic plasticity.
    \item \textbf{Mechanistic vs. Integrative Approach:} Previous STDP models often adopt a mechanistic perspective, focusing on the molecular mechanisms that govern synaptic changes. However, these models tend to overlook the comprehensive and integrative nature of consciousness. CD-STDP bridges this gap by incorporating insights from cognitive neuroscience and consciousness studies, recognizing consciousness as an emergent property of complex neural networks.
    \item \textbf{Unconscious vs. Conscious Processing:} Traditional STDP models primarily address unconscious neural processes involved in learning and memory. They often neglect the potential role of consciousness in shaping synaptic plasticity. CD-STDP, on the other hand, highlights the significance of conscious processing, suggesting that conscious awareness can modulate neural activity patterns and, consequently, synaptic changes.
    \item \textbf{From Correlation to Causation:} While traditional STDP models focus on the correlation between neuronal firing patterns and synaptic modifications, CD-STDP delves deeper into the causal relationship between consciousness and synaptic plasticity. It proposes that conscious experiences actively influence the dynamics of neural circuits, leading to specific patterns of synaptic strengthening or weakening.
\end{enumerate}

Overall, the transition from traditional STDP models to CD-STDP represents a exemplar shift in learning of SNNs, acknowledging the inseparable link between consciousness and neural function and serves to bridge the gap between the mechanistic understanding of synaptic plasticity and the subjective experience of consciousness. This set out aims to deepen the comprehension to explore the role of synaptic plasticity in higher cognitive functions, address the hard problem of consciousness, inform artificial intelligence and neuromorphic engineering, and other relevant disciplines. By infusing consciousness into STDP, this article attempt to develop a more comprehensive understanding of how synaptic plasticity interacts with subjective experience, potentially leading to profound insights into the nature of consciousness along with SNNs.

This article aims to provide:
\begin{enumerate}
    \item Proposed a consciousness based learning STDP model i.e. $CD$-STDP.
    \item Implementation of $CD$-STDP with SNNs on image classification task.
    \item Analysis of $CD$-STDP with different metrics along with the properties of consciousness.
\end{enumerate}
The rest of the  article is organized as follows: In Section \rom{2} the article includes the fundamental of consciousness and STDP to establish the groundwork for our study. In Section \rom{3} the article includes the design and construction of $CD$-STDP. In Section \rom{4}, the article includes the methodology for the implementation of proposed model in network. In Section \rom{5}, the proposed model is thoroughly evaluated for the image classification task and its conscious parameters are being analysed. Finally, Section \rom{6}, concludes the article.
  
\section{FUNDAMENTALS OF Consciousness AND STDP}
\subsection{Basics of Consciousness}
Consciousness, a fundamental aspect of human experience, includes key properties like, subjectivity (wherein individual awareness is intrinsic and inaccessible to others), qualia (the subjective qualities defining experiences), intentionality (directing consciousness towards objects or states), unity (integrating diverse stimuli into a cohesive experience), temporality (manifesting as a continuous flow of evolving experiences), self-awareness (enabling reflection on one's identity and mental states), adaptiveness (serving functions conducive to survival, varying levels from deep sleep to heightened awareness, and neural correlates, hinting at the biological underpinnings) \cite{mandler2002consciousness}. Despite extensive exploration across disciplines, understanding consciousness remains a profound challenge, with diverse theories seeking to explain its nature from neuroscientific accounts to Artificial Intelligence perspectives.
\subsection{Integrated Information as a Metric or Phi}

IIT provides a framework for understanding consciousness. The centre of the idea is "Phi" ($\phi$), which quantifies the amount of integrated information generated by a system. Integrated information refers to the extent to which the system's components interact in a manner that is irreducible to the interactions of its parts taken separately \cite{tononi2012integrated}.

Mathematically, Phi can be defined as follows:

Let's consider a system with N elements. Each element can exist in one of M states. The state of the entire system can be represented by a vector, s, of length $M^N$, where each entry represents a possible configuration of the system's elements. For any given system state $s_i$, we can calculate the $\phi$ value as follows:

\textbf{Cause-Effect Repertoire (CER):} This represents all the possible past and future states of the system that can be caused by the current state, $s_i$. It's denoted as CER($s_i$).

\textbf{Cause-Effect Information (CEI):} This measures the amount of information generated by the system. It's defined as the Kullback-Leibler divergence between the system's probability transition matrix T and the product of its marginal probability distributions. Mathematically, CEI is expressed as:

\begin{equation}
    CEI(s_i) = D_{(KL)}(T||T^{(1)}*T^{(2)}*...*T^{(n)})
\end{equation}

where 
$D_{(KL)}$ is the Kullback-Leibler divergence, T is the transition matrix of the system, and 
$T^{(k)}$ is the marginal transition matrix over the k-th element.

Integrated Information (Phi): Phi quantifies the integrated information generated by a system, capturing the extent to which the system's parts interact in a way that's irreducible. It is defined as the difference between the cause-effect information generated by the whole system and the sum of the cause-effect information generated by its parts:
\begin{equation}
    \phi(s_i)=  CEI(s_i)-\sum_{k}CEI(s_i^{(k)}) 
\end{equation}

where $s_i^{(k)}$ denotes the state of the system with the k-th element constrained to one of its possible states.

Higher $\phi$ values indicate a higher level of integrated information and, according to IIT, are associated with a higher level of consciousness. However, calculating $\phi$ for complex systems can be computationally challenging due to the vast number of possible states and interactions, especially for large-scale neural systems.

\subsection{Model of Spiking Neuron and Synaptic Plasticity in Computational Systems}
The core computing unit in SNNs is the spiking neuron, influenced by input information through plastic synapses \cite{ahmed2020brain}. The Leaky Integrate-and-Fire (LIF) model, widely used in deep learning, efficiently captures biological neuron dynamics with simple discrete equations \cite{yamazaki2022spiking}. The LIF model is represented by:
\begin{equation}
    \tau_{mem}\frac{dV_{mem}}{dt}= -V_{mem}+ W*\delta (t-t_i)
\end{equation}
where
$V_{mem}$ is the membrane potential, $W$ is the synaptic weight, $t_i$ is the time of input spike occurrence, $\tau_{mem}$ is the membrane time constant. In this model, input dynamics are expressed through dirac delta spike trains, denoted as $\delta(t - t_i)$, influenced by synaptic weights. The neuron integrates input current, causing its membrane potential to rise. Once input spikes crosses a predefined threshold potential, an output spike is emitted and the potential undergoes exponential decay, after which the membrane potential resets. \cite{yadav2023comparative}.

Biological synapses display intricate dynamics that regulate how the timing patterns of incoming spike trains influence the resulting pattern of postsynaptic responses. Similarly, in SNNs, the strength of synapses between pre- and post-neurons adjusts through STDP \cite{lee2018training}. According to STDP theory, synaptic strength undergoes exponential adjustments based on the timing correlation observed in the spiking patterns of the involved neurons. This dynamic is mathematically represented as:

\begin{equation}
    \Delta W = \left\{\begin{matrix}
A^+ e^{\frac{-|t_{pre}-t_{post}|}{\tau_+}}&; t_{pre}\leq t_{post} \\ 
-A^-  e^{\frac{-|t_{pre}-t_{post}|}{\tau_-}}& ;t_{pre}>  t_{post} 
\end{matrix}\right.
\end{equation}

Here, $\Delta W$ represents the change in synaptic strength, $A^+$ and $A^-$ denote the coefficients of LTP and LTD respectively, and $\tau_+$ and $\tau_-$ represent the time constants associated with LTP and LTD. Fig. 1 illustrates how alterations in synaptic strength ($\Delta W$) correlate with the timing parameters $t_{\text{pre}}$ and $t_{\text{post}}$ governed by STDP.

\begin{figure}
    \centering
    \includegraphics[width = 8cm, height = 6.5cm]{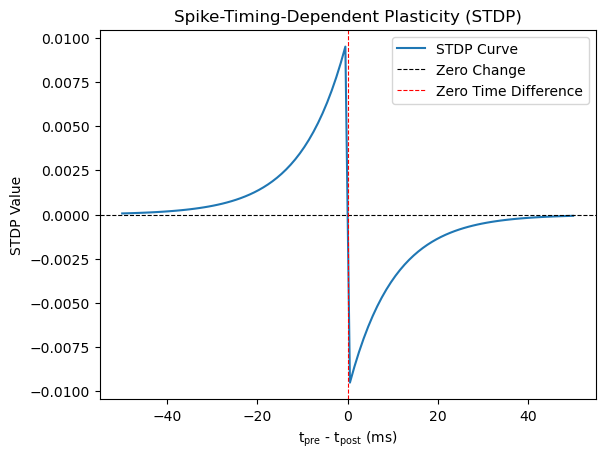}
    \caption{{Synaptic weight value based on time difference. }}
    \label{fig:enter-label}
\end{figure}

\section{Proposed CD-STDP Design}
\subsection{Theoretical Hypothesis}
STDP and Consciousness emerges from the dynamic activity of neural network modulation. This work like learning and adaptation as, STDP facilitates the learning and adaptation of neural networks to incoming sensory information by allowing the network to form associations and extract meaningful information from the neuronal synchrony \cite{griffith2022spike}. While consciousness is thought to involve the integration of information across distributed neural network \cite{tononi2010information}. STDP could contribute to this integration of information by promoting synchrony among neurons that are involved in processing related information. Synchronous firing of neurons strengthens their connections, leading to the formation of functional networks that underlie conscious experiences \cite{gansel2022neural}. This can also perform as selective attention and awareness as STDP also play a role in selective attention, the process by which the neural networks prioritizes certain stimuli for further processing. Also the temporal dynamics of consciousness and STDP could contribute to the subjective experience of time and the perception of temporal order in conscious awareness \cite{buonomano2017your}. Changes in synaptic strength over time, driven by STDP, could influence the flow of information processing and contribute to the sense of continuity in conscious experience.

\subsection{Assumptions}
\begin{itemize}
    \item The effect of consciousness came from the coefficient of LTP and LTD during learning. 
    \item The length of pre-synaptic and post-synaptic spike trains are equal and fixed.
    \item The amplitude of all the spikes is considered constant (Amplitude = 1).
    \item The correlation between individual spikes are short-lived, meaning the influence of a spike on the synaptic weight decays quickly as we move away from the precise timing of the spike.
    \item The decay of correlation with time difference ($\Delta t$) follows an exponential decay.
    
\end{itemize}
\subsection{Dependency factors in STDP for conciousness}
In STDP, which is a synaptic plasticity rule governing the strengthening or weakening of connections between neurons based on the relative timing of their spikes while consciousness is a complex phenomenon that involves the coordinated activity of many neurons in the network. While STDP play a role in shaping neural networks involved in consciousness, it's important to note that consciousness itself is not fully understood, and its precise neural correlates are still being investigated. The dependency factors of consciousness in the context of STDP, can be look through various aspects such as:

\textbf{Network Connectivity:} The connectivity patterns within neural network could influence the emergence of conscious states. STDP could play a role in shaping these networks by strengthening or weakening specific synaptic connections based on their timing and activity.

\textbf{Spike Timing}: The precise timing of neuronal spikes is crucial for information processing in the brain. STDP mechanisms could be involved in coordinating the timing of neuronal activity within specific circuits that contribute to conscious processing.

\textbf{Neuronal Synchrony:} Synchronous activity among neurons is important for certain aspects of consciousness, such as the integration of information across neural network. STDP can influence the synchronization of neuronal activity by shaping the strength of synaptic connections involved in synchronization mechanisms.

If all the above aspects are considered then it can be clearly seen that consciousness is closely related to learning and memory processes. STDP is a form of synaptic plasticity that underlies learning and link associated between neurons, and it could contribute to the formation of neuronal representations and networks that support conscious experiences.

\subsection{Construction of  Conscious Coefficient of LTP and LTD}
First, let's define $H(X_{\text{M}0} | X_{\text{M}1})$ as the conditional entropy of past state of the hidden layer $X_{\text{M}0}$ given the current state $X_{\text{M}1}$ and $H(X_{\text{M}1} | X_{\text{M}0})$  as the conditional entropy of the given states of the hidden layer $X_{\text{M}1}$ given the past state $X_{\text{M}0}$.

Now, to make $\phi_{\text{pos}}$ and $\phi_{\text{neg}}$ the conscious coefficients, the concept of $\phi$ from IIT theory is taken.
To derive expressions for these quantities consider $A^+$ = $\phi_{\text{pos}}$, which measures the reduction in uncertainty about the past states of the hidden layer ($X_{\text{M}0}$) given the current states ($X_{\text{M}1}$), we can express it as:

\begin{equation}
\phi_{\text{pos}} = H(X_{\text{M}0}) - H(X_{\text{M}0} | X_{\text{M}1})
\end{equation}

Similarly, consider $A^-$=$\phi_{\text{neg}}$, which measures the reduction in uncertainty about the current states of the hidden layer ($X_{\text{M}1}$) given the past states ($X_{\text{M}0}$), we have:

\begin{equation}
\phi_{\text{neg}} = H(X_{\text{M}1}) - H(X_{\text{M}1} | X_{\text{M}0})
\end{equation}

Now, to derive expressions for the conditional entropies $H(X_{\text{M}0} | X_{\text{M}1})$ and $H(X_{\text{M}1} | X_{\text{M}0})$.

Using the definition of conditional entropy, we have:

\begin{equation}
H(X_{\text{M}0} | X_{\text{M}1}) = - \sum_{x_{\text{M}0}, x_{\text{M}1}} P(x_{\text{M}0}, x_{\text{M}1}) \log P(x_{\text{M}0} | x_{\text{M}1})
\end{equation}

\begin{equation}
H(X_{\text{M}1} | X_{\text{M}0}) = - \sum_{x_{\text{M}0}, x_{\text{M}1}} P(x_{\text{M}0}, x_{\text{M}1}) \log P(x_{\text{M}1} | x_{\text{M}0})
\end{equation}

These expressions represent the uncertainty associated with predicting $X_{\text{M}0}$ given $X_{\text{M}1}$, and vice versa.

These expressions are used to compute $\phi_{\text{pos}}$ and $\phi_{\text{neg}}$ based on the specific distributions of $X_{\text{M}0}$ and $X_{\text{M}1}$ in the neural network model. Also the present state refers to the current firing activity of neurons in the network. It represents the pattern of spikes occurring at the current moment or within a short time window around the current time. The present state encapsulates the ongoing activity and dynamics of the network as it processes input stimuli or performs computations.
Similarly, the past state refers to the firing activity of neurons in the network at previous time points relative to the current moment. It represents the history of spike events and the temporal context in which the network operates. The past state includes information about the recent activity patterns of neurons leading up to the current moment.

\subsection{Construction of $CD$-STDP}
In the STDP model, the synaptic strength is characterized by weights, and the alteration of the synaptic strength is influenced by the time difference between pre-synaptic spike moment ($t_{pre}$) and post-synaptic spike moment ($t_{post}$). However in the proposed STDP the alteration is influenced by the conscious coefficients of LTP and LTD proposed in equation (5,6) along with the time difference between $t_{pre}$ and $t_{post}$. This relation can be represented by Equation (9).

\begin{equation}
    \Delta W = \left\{\begin{matrix}
\phi_{pos} e^{\frac{-|t_{pre}-t_{post}|}{\tau_+}}&; \Delta t \leq 0 \\ 
\phi_{neg}  e^{\frac{-|t_{pre}-t_{post}|}{\tau_-}}& ;\Delta t > 0 
\end{matrix}\right.
\end{equation}

with $\tau_+$ and $\tau_-$ be the time constants of LTP and LTD respectively.
to prevent weights from falling below zero or from exploding an auxiliary clauses is needed, for that define $W_{max}$ and $W_{min} = 0$. If $W_{old} + \Delta W > W_{max}, W_{new} \rightarrow W_{max}$; if $W_{old} + \Delta W < 0, W_{new} \rightarrow 0$. With that STDP changes the value $W_{new} = W_{old} + \Delta W$ according to:\\
\begin{equation}
    W_{new} = \left\{\begin{matrix}
min\{{W_{max}, W_{old} +\phi_{pos}e^{\frac{-|t_{pre}-t_{post}|}{\tau_+}}\}}&; \Delta t \leq 0 \\ 
max\{{0, W_{old} -\phi_{neg}e^{\frac{-|t_{pre}-t_{post}|}{\tau_-}}\}}& ; \Delta t > 0
\end{matrix}\right.
\end{equation}

An important part of $CD$-STDP rule is the scheme of pairing pre-and post-synaptic spikes and taking the effects of dynamic LTP and LTD in form of conscious elements $\phi_{pos}$ and $\phi_{neg}$ when evaluating weight change.\par 

The algorithm for implementation of the proposed model is described in Algorithm 1.

\begin{algorithm}
  \caption{$CD$-STDP Algorithm}
  \begin{algorithmic}[1]
      \\
    \textbf{Input:}\\
    \hspace{0.5cm} $t_{pre},\ t_{post},\ w_{old}$\\
    \vspace{0.1cm}
    \textbf{Initialize:}\\
    \hspace{0.5cm} $\tau_+$, $\tau_-$, $T$,\ $W_{max}$, \ $X_{\text{M}0}$\\
    \textbf{Working:}    
    \For{$t_{pre}, t_{post} \in [1, T]$}
       \State $\Delta t = t_{pre} - t_{post}$ 
        \If{$\Delta t \leq 0$}
           \State $\phi_{\text{pos}} = H(X_{\text{M}0}) - H(X_{\text{M}0} | X_{\text{M}1})$
           \State $\Delta W =\phi_{\text{pos}} * e^{-\frac{|\Delta t|}{\tau_+}}$
        \Else 
           \State $\phi_{\text{neg}} = H(X_{\text{M}1}) - H(X_{\text{M}1} | X_{\text{M}0})$
           \State $\Delta W = \phi_{\text{neg}} * e^{-\frac{|\Delta t|}{\tau_-}}$
        \EndIf

        \If{$0 \leq W_{old} + \Delta W \leq W_{max}$}
            \State \Return $W_{new} = W_{old} + \Delta W$
        \ElsIf{$ W_{old} + \Delta W > W_{max}$}
            \State \Return $W_{new} = W_{max}$
        \ElsIf{$ W_{old} + \Delta W < W_{max}$}
            \State \Return $W_{new} = 0$ 
        \EndIf            
    \EndFor    

  \end{algorithmic}
\end{algorithm}

\section{Methods}
\subsection{Encoding, Network Topology and Learning}
In SNN the first step is encoding. Poisson-distributed rate coding is used as the preferred method for transforming images into binary spikes, surpass alternatives like temporal, phase, and burst coding schemes \cite{zhang2019tdsnn}. This approach generates spike trains over multiple time steps, with spike count proportional to pixel intensity. Practically, it involves  pixel's value  within the range [0, 255] to generated spike train using poisson distribution. These spike trains then feed into a SNN. Within the network architecture, neurons in the hidden layer are used as mathematical models defined by Equation (1), with synaptic connections playing a critical role. Synaptic connections link neurons, transmitting spikes and influencing behavior based on the learning algorithm employed. The network's topology used in the work is a fully connected feedforward arrangement, as it performs better as compared to other topologies like  sparsely connected or structured arrangements \cite{sammut2011encyclopedia}. Plasticity mechanisms, i.e. STDP, drive synaptic adaptability, crucial for tasks like classification and pattern recognition, by learning temporal relationships between neuron firings ($t_{pre}$ and $t_{post}$). During training, the proposed $CD$-STDP model adjusts synaptic weights based on spike timing differences, using conscious elements like $\phi_{\text{pos}}$ and $\phi_{\text{neg}}$ as described in Equations (5,6), thereby enhancing the network's ability to classify and recognize patterns. The working for implementation of the $CD$-STDP in SNN is described in Algorithm 2.

\begin{algorithm}
  \caption{SNN working with $CD$-STDP}
  \begin{algorithmic}[1]
  
    \\
    \textbf{Input:}\\
    \hspace{0.5cm} Image $X$\\
    \vspace{0.1cm}
    \textbf{Output:} \\ 
    \hspace{0.5cm} Output spike train $t_{post}$
    \vspace{0.1cm}
    \State \textbf{Parameters:} \\
    \hspace{0.5cm} $n_0$= size of image, \\
    \hspace{0.5cm}$n^i$= number of neurons in $i^{th}$ layer,\\
    \hspace{0.5cm} $T_{max}$= total simulation time,\\
    \hspace{0.5cm} $t_{pre}$= input spike train,\\
    \hspace{0.5cm} $dt$= time steps\\
    \vspace{0.1cm}  
    \textbf{Initialize:}
    \hspace{0.5cm} \textbf{for} i = 1 to $n^i$ \textbf{do}\\
    \hspace{1cm} $V_i$ = 0\\
    \hspace{0.5cm} \textbf{end for}\\
    
    \vspace{0.1cm}    
    \textbf{Phase 1: Input Encoding} \\
    \hspace{0.5cm} \textbf{for} i = 1 to $n^i$ \textbf{do}\\
    \hspace{1cm} $\lambda$ = $x_i$/$\tau$\\
    \hspace{1cm} $t_{pre}$ = Poisson spike ($\lambda$, dt)\\
    \hspace{0.5cm} \textbf{end for}\\
    
    \vspace{0.1cm}    
    \textbf{Phase 2: Forward Pass} \\
    \hspace{0.5cm} \textbf{for} dt = 1 to $t_{max}$ \textbf{do}\\
    \hspace{1cm} $I(t)$ $\xleftarrow{}$ Integrate ($W^t$, $t_{pre}$)\\
    \hspace{1cm} $V_m(t)$ $\xleftarrow{}$ Accumulate ($V(t-1)$, $I(t)$)\\
    \hspace{1cm} \textbf{if} $V_m(t)$ $>$ $V_{th}$ \textbf{then} \\
    \hspace{1.5cm} $t$ $\in$ $t_{post}$\\
    \hspace{1cm} \textbf{end if}\\
    \hspace{0.5cm} \textbf{end for}\\

    \vspace{0.1cm}    
    \textbf{Phase 3: Learning} \\ 
    \hspace{0.5cm} As in Algorithm 1
    
  \end{algorithmic}
\end{algorithm}

\subsection{Granger Causality}
In Neuroscience, Granger causality has been used in analyzing causal relationships among various brain regions and their functional connectivity. For instance, research has shown that Granger causality can effectively perceive the direction of information flow between different brain regions during cognitive tasks \cite{dhamala2008analyzing}. It's a method employed to examine the causal relationship between two sets of time series data. In this article, granger causality is used to find the causal relationship between input spikes and the output spikes. Let's denote the input spike train as $X(t)$ and the output spike train as $Y(t)$, where \textit{t} represents time. Granger causality tests whether the pre-synaptic spike train $X(t)$ Granger-causes the post-synaptic spike train $Y(t)$ by assessing the predictive power of past values of $X(t)$ in explaining the current values of $Y(t)$, after accounting for the past values of $Y(t)$. Here Granger causality is implemented using a Vector Autoregressive (VAR) model. In this model, each time $X(t)$ and $Y(t)$ is regressed on its own past values as well as the past values of the other spike train. Mathematically, the VAR model for two spike trains can be represented as follows:
\begin{equation}
    X(t) = \sum_{i=1}^{p} a_{1i}X(t-i)+\sum_{i=1}^{p} b_{1i}Y(t-i)+\epsilon _{1t}
\end{equation}
\begin{equation}
  Y(t) = \sum_{i=1}^{p} a_{2i}X(t-i)+\sum_{i=1}^{p} b_{2i}Y(t-i)+\epsilon _{2t}  
\end{equation}

Here $a_{1i}$ and $a_{2i}$ are coefficients representing the influence of past values of $X(t)$ on $X(t)$ and $Y(t)$ respectively, $b_{1i}$ and $b_{2i}$ are coefficients representing the influence of past values of $Y(t)$ on $X(t)$ and $Y(t)$ respectively, $p$ is the order of the VAR model, and $\epsilon_{1t}$ and $\epsilon_{2t}$ are error terms.
 The Granger causality metric is computed using the F-test statistic, which compares the fit of a VAR model. The F-test statistic for Granger causality can be calculated as follows:
 \begin{equation}
     F = \frac{(RSS_{reduced}-RSS_{full})/m}{RSS_{full}/(n-2p-1)}
 \end{equation}

 where $RSS_{reduced}$ is the residual sum of squares from the VAR model without the inclusion of past values of the pre-synaptic spike train, $RSS_{full}$ is the residual sum of squares from the VAR model with the inclusion of past values of the pre-synaptic spike train, \textit{m} is the number of restrictions (the difference in the number of parameters between the reduced and full models), and \textit{n} is the number of observations (total number of time points).
 \begin{figure}
     \centering
     \includegraphics[width= 8.85cm,height= 8cm]{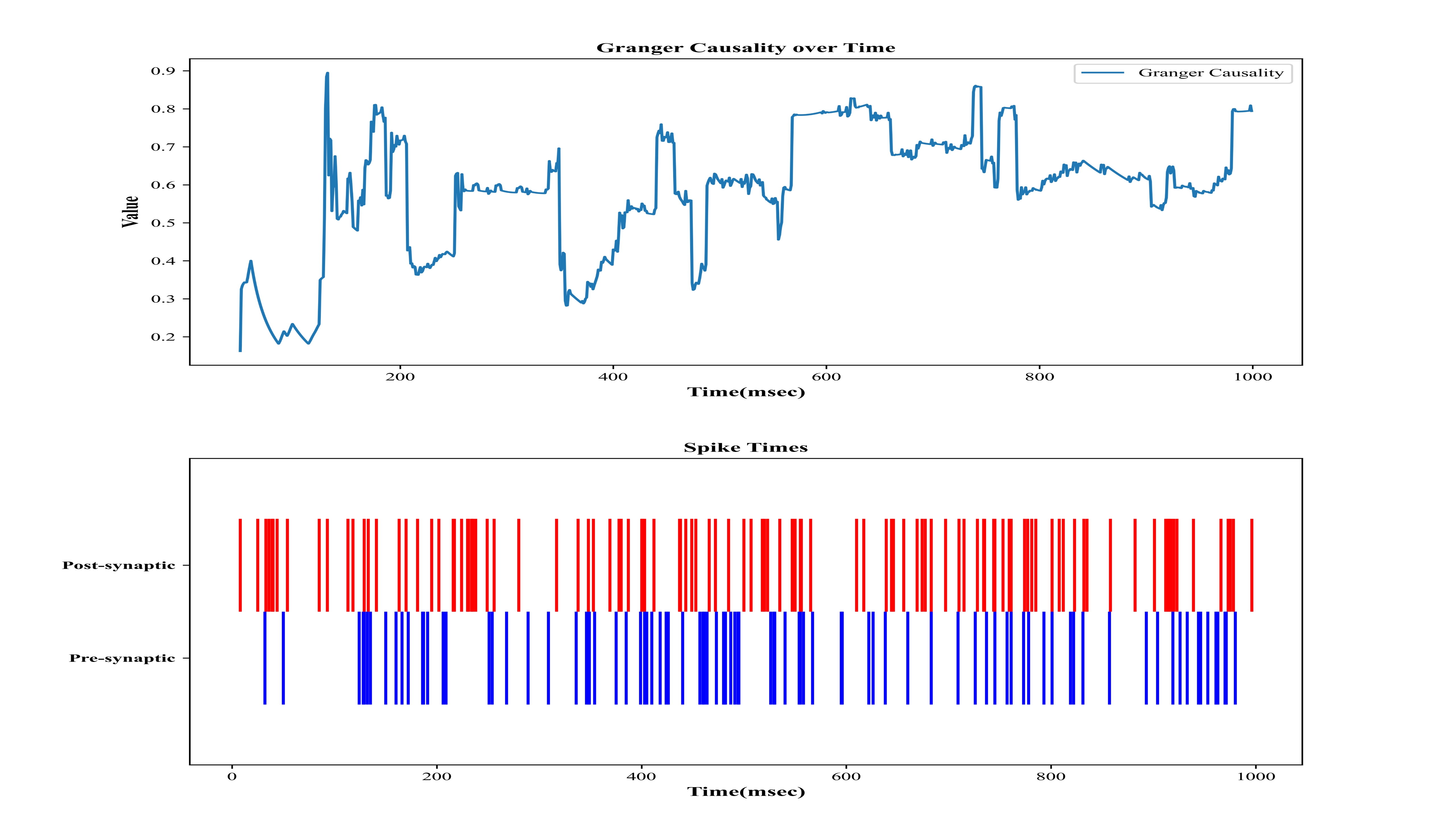}
     \caption{Granger causality for two neurons}
     \label{fig:enter-label}
 \end{figure}

The value of F-test statistic for Granger causality for pre-synaptic and post-synaptic spike train of length 1 second over two neurons is represented in Fig. 2.

\section{Results and Discussion}
The current study is based on the design and implementation of the proposed $CD$-STDP model for image recognition task. The design part is presented in Section \rom{3} of the article. In this section experiment or implementation of the proposed model is being discussed. The image classification experiment is performed on MNIST, FashionMNIST and CALTECH (Face/Motorbike) datasets, which are commonly used for benchmarking SNN learning algorithms \cite{lee2018deep}. The MNIST and FashionMNIST datasets  comprises 60,000 grayscale images for training and 10,000 for testing. Featuring single digits ranging from 0 to 9 in case of MNIST and items of 10 type of different clothing in FashionMNIST, each 28x28 in dimension. The CALTECH (Face/Motorbike) dataset used is a subset of 1226 images from the Caltech dataset spanning two different object categories, namely, Face and Motorbike, each 300x200 in dimension. To implement model on these datasets, in contrast to more complex architectures, the authors opted for a streamlined approach by employing a single hidden layer in algorithm. This decision was grounded in the pursuit of simplicity and efficiency, as it allows for a more transparent understanding of the model’s inner workings and reduces computational overhead. The choice of a single hidden layer is strategic, as it facilitates a straightforward interpretation of the model’s behavior and also aligns with the goal of showcasing the inherent strength and adaptability of the proposed algorithm. Datasets are divided into mini-batches for training that achieves learning with fewer synaptic weight updates. PyTorch library along with snnTorch library \cite{eshraghian2023training} is used for all the experiments. This choice is motivated by its robust support for accelerated and memory efficient training \cite{yang2023lc}. The authors adopt a discrete-time simulation approach for SNNs, utilizing a time step of 1 milliseconds.
Each layer undergoes simulation over 100 discrete time steps. The remaining parameters are presented in Table I.

\begin{table}[]
\centering
\caption{\footnotesize{$A$-STDP in SNN SIMULATION PARAMETERS}}
\begin{tabular}{p{4cm} p{1cm}}
\hline 
\textbf{Parameter}                             & \textbf{Values} \\ \hline \hline
Synaptic Weight Range                          & {[}0,1{]}       \\ 
$\tau_+$, $\tau_-$ & 0.5             \\ 
$\tau_{mem}$                   & 10 msec         \\ 
$\gamma$                   & 0.70         \\
$\eta$                   & 0.20         \\ 
Threshold Potential                            & 1               \\ 
Mini-batch size                                & 128             \\ 
Resting potential                              & 0.0 mV          \\ 
Reset potential                                & 0.0 mV          \\ \hline \hline
\end{tabular}
\end{table}

Table \rom{2} reports the results of the purposed model on MNIST dataset. The proposed model achieved the classification accuracy of 98.6\%. For the validation of the accuracy of the proposed model, the training accuracy curve along with the change in values of $\phi_{\text{pos}}$ and $\phi_{\text{neg}}$ is presented in Fig. 3. Table \rom{3} reports the results of the purposed model on FashionMNIST dataset. The proposed model achieved the classification accuracy of 85.61\% and Table \rom{4} reports the results of the purposed model on CALTECH (FACE/MOTORBIKE) dataset. The proposed model achieved the classification accuracy of 99.0\%.  The efficacy of the proposed algorithm is achieving a good accuracy while being in compact in size due to single hidden layer. Notably, as presented in the last row of the Table \rom{2}, Table \rom{3} and Table \rom{4}, the proposed model achieves competitive classification accuracy across different datasets with most compact structure in comparison with typically fully connected SNNs.
In order to effectively benchmark and underscore the efficacy of the proposed algorithm, a comprehensive comparative analysis was conducted against existing works in the field.
This comparative study serves the dual purpose of evaluating the performance of the proposed algorithm and highlighting its potential in relation to established methodologies. The
proposed $CD$-STDP demonstrates competitive classification accuracy across various datasets while employing fewer parameters. For example in methodology used in this article for
classification uses only single layer and also doesn’t use any convolution layer which highly reduce the number of trainable parameters.

\begin{table*}[]
\centering
\caption{\footnotesize{CLASSIFICATION ACCURACY OF $CD$-STDP FOR MNIST.}}
\begin{tabular}{p{4cm}p{4cm}p{4cm}p{3cm}}
\hline
Model               & Learning Rule               & Type             & Accuracy \\ \hline \hline
\cite{querlioz2013immunity}      & STDP                           & unsupervised     & 93.5\%    \\ 
\cite{diehl2015fast} & STDP                           & Unsupervised     & 95.0\%    \\ 
\cite{hao2020biologically}         & Sym-STDP+SVM                   & Un- \& Supervised & 96.7\%    \\ 
DCSNN \cite{mozafari2019bio}        & D0G+STDP+R-STDP                & Un- \& Supervised & 97.2\%    \\ 
\cite{falez2019multi}       & DoG+STDP+SVM                   & Un- \& Supervised & 98.6\%    \\ 
SDNN \cite{kheradpisheh2018stdp}          & DoG+STDP+SVM                   & Un- \& Supervised & 98.4\%    \\ 
SpiCNN \cite{lee2018deep}             & LoG+STDP                       & Un- \& Supervised & 91.1\%    \\ 
\cite{tavanaei2017multi}     & STDP+SVM                       & Un- \& Supervised & 98.4\%    \\ 
\cite{ferre2018unsupervised}         & STDP+BP                        & Un- \& Supervised & 98.5\%    \\
SSTDP \cite{li2022bsnn}          & STDP+BP                        & Supervised       & 98.1\%    \\ 
VPSNN \cite{zhang2018plasticity}          & Equilibrium Propagation + STDP & Supervised       & 98.5\%    \\ 
CBSNN \cite{shi2020curiosity}         & VPSNN+Curiosity                & Supervised       & 98.6\%    \\ 
BP-STDP \cite{tavanaei2019bp}       & STDP-Based BP                  & Supervised       & 97.2\%    \\ 
GLSNN \cite{zhao2020glsnn}         & Global Feedback +STDP          & Supervised       & 98.6\%    \\ 
\cite{dong2023unsupervised}         & ASF+ATB+ALIC+STB-STDP          & Unsupervised     & 97.9\%    \\
\cite{cai2023spike}                & BSTDG                        & Supervised            & 98.7\%      \\ 
\cite{tavanaei2016bio}               & Sparse Coding+STDP+SVM                       &  Unsupervised              & 98.3\%    \\ 
Proposed work           & CD-STDP                        & Unsupervised     & 98.6\%    \\ \hline
\end{tabular}
\end{table*}

\begin{figure}
    \centering
    \includegraphics[width= 10.85cm,height= 9cm]{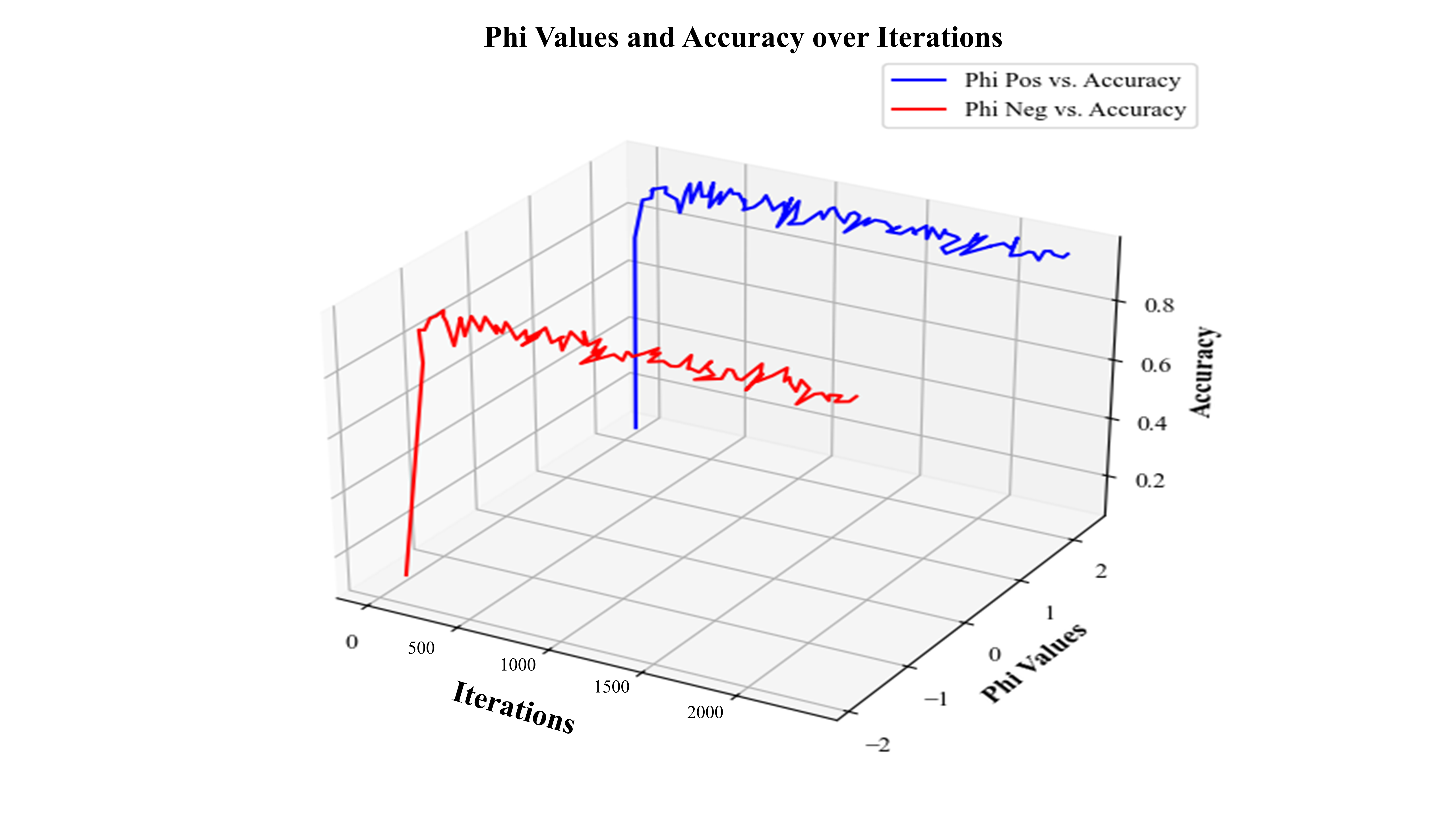}
    \caption{Accuracy vs Change in value of phi for MNIST dataset}
    \label{fig:enter-label}
\end{figure}

\subsection{Comparison with Related Works }
\begin{table*}[]
\centering
\caption{\footnotesize{CLASSIFICATION ACCURACY OF $CD$-STDP FOR FashionMNIST.}}
\begin{tabular}{p{4cm}p{4cm}p{4cm}p{3cm}}
\hline
Model              & Learning Rule              & Type         & Accuracy \\ \hline \hline
VPSNN \cite{zhang2018plasticity}       & Equilibrium Propagation+STDP & Supervised   & 83.0\%   \\ 
FSpiNN \cite{putra2020fspinn} & STDP                         & Unsupervised & 68.8\%   \\ 
\cite{hao2020biologically}         & Sym-STDP                     & Supervised   & 85.3\%   \\ 
CBSNN \cite{shi2020curiosity}     & VPSNN+Curiosity              & Supervised   & 85.7\%   \\ 
GLSNN \cite{zhao2020glsnn}   & Global Feedback+STDP         & Supervised   & 89.1\%   \\ 
\cite{rastogi2021self}       & A-STDP                       & Unsupervised & 75.9\%   \\ 
\cite{dong2023unsupervised}       & ASF+ATB+ALIC+STB+STDP        & Umsupervised & 87.0\%   \\ 
Proposed work         & CD-STDP                      & Unsupervised & 85.61\% \\ \hline \hline
\end{tabular}
\end{table*}

\begin{table*}[]
\centering
\caption{\footnotesize{CLASSIFICATION ACCURACY OF $CD$-STDP FOR CALTECH (FACE/MOTORBIKE).}}
\begin{tabular}{p{4cm}p{4cm}p{4cm}p{3cm}}
\hline
Model & Learning Rule  & Type     &  Accuracy \\ \hline \hline
 SDNN   \cite{kheradpisheh2018stdp}               & STDP+SVM         &Un- \& Supervised           & 99.1\%            \\ 
Spiking CNN  \cite{masquelier2007unsupervised}              &  STDP+RBF      & Unsupervised           & 97.7\%            \\ 
Spiking CNN  \cite{mozafari2018first}     &  Reinforcement STDP       &  Supervised          & 98.9\%            \\ 
SpiCNN    \cite{lee2018deep}                  &   LoG+STDP       &  Un- \& Supervised & 97.6\%            \\ 
\cite{liu2021sstdp}                 & SSTDP             & Supervised            & 99.3\%            \\ 
 \cite{cai2023spike}                 & BSTDG             & Supervised                & 99.0\%            \\ 
Proposed work              & $CD$-STDP           & Unsupervised  & 99.0\%            \\ \hline \hline
\end{tabular}
\end{table*}

Previous research attempts have investigated the application of STDP in training SNNs. In \cite{masquelier2007unsupervised}, the feasibility of utilizing STDP is demonstrated by integrating it with an ANN-based Radial Basis Function, incorporating a single convolutional layer in an SNN. Additionally, \cite{kheradpisheh2018stdp, tavanaei2017multi}, successful training of SNNs is achieved by combining STDP with an ANN-based classifier, specifically using the Support Vector Machine, and incorporating multiple convolutional layers. Both studies, \cite{masquelier2007unsupervised} and \cite{kheradpisheh2018stdp}, adopted a temporal rank-order spike encoding scheme within a network composed of integrate-and-fire neurons. They trained the convolutional layers either by using STDP+RBF or with STDP+SVM respectively to improve the efficacy of the network. On the other hand, both \cite{cai2023spike} and \cite{liu2021sstdp} integrate backpropagation with STDP in their own ways to improve the efficacy of the network. Also, this work is different form the work described in \cite{tavanaei2016bio, mozafari2018first }, where they use reward based system, constant scaling parameters or sparse coding with STDP respectively. In \cite{lee2018deep, falez2019multi, mozafari2019bio,hao2020biologically}, researchers used filters like DoG, LoG alongside STDP to achieve the desired accuracy. However, in \cite{ferre2018unsupervised, li2022bsnn, zhang2018plasticity, shi2020curiosity, tavanaei2019bp, zhao2020glsnn}, researchers uses a supervised mix method along with STDP to improve the efficacy of the network which is similar complex as of the traditional ANN as, the researchers uses either backpropogation, curiosity or Global Feedback mechanism. Else, \cite{querlioz2013immunity, diehl2015fast, putra2020fspinn, rastogi2021self} uses either simple STDP or $A$-STDP to train the network but can't get much remarkable accuracy on the standard datasets. While the proposed model using only CD-STDP as learning algorithm provides a competitive accuracy on standard datasets.

\subsection{Analysis of Conscious Elements and Consciousness of the Network}
\begin{figure}
    \centering
    \includegraphics[width= 8.85cm,height= 5cm]{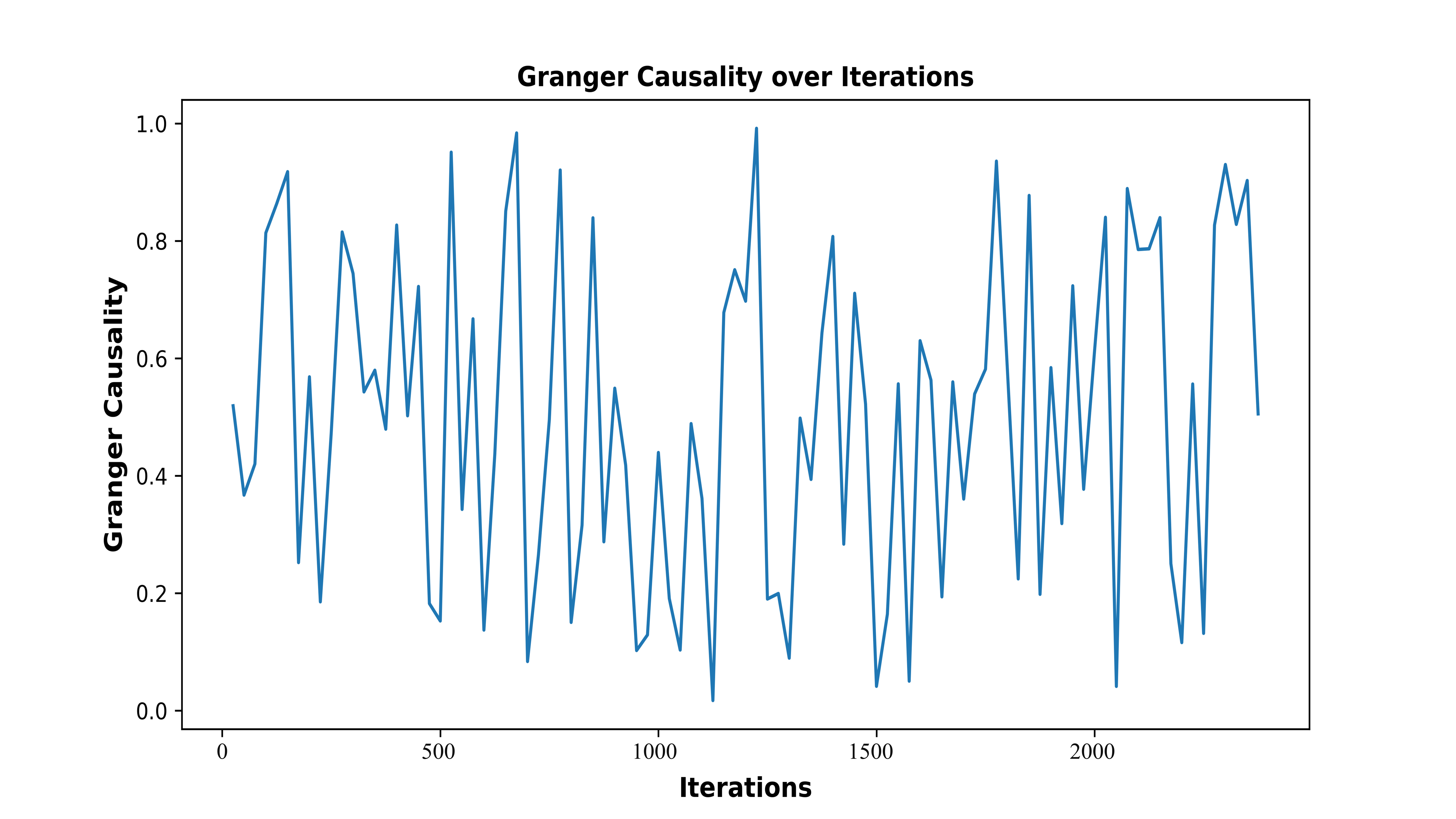}
    \caption{Granger causality for the network on MNIST dataset}
    \label{fig:enter-label}
\end{figure}

\begin{figure}
    \centering
    \includegraphics[width= 8.85cm,height= 5cm]{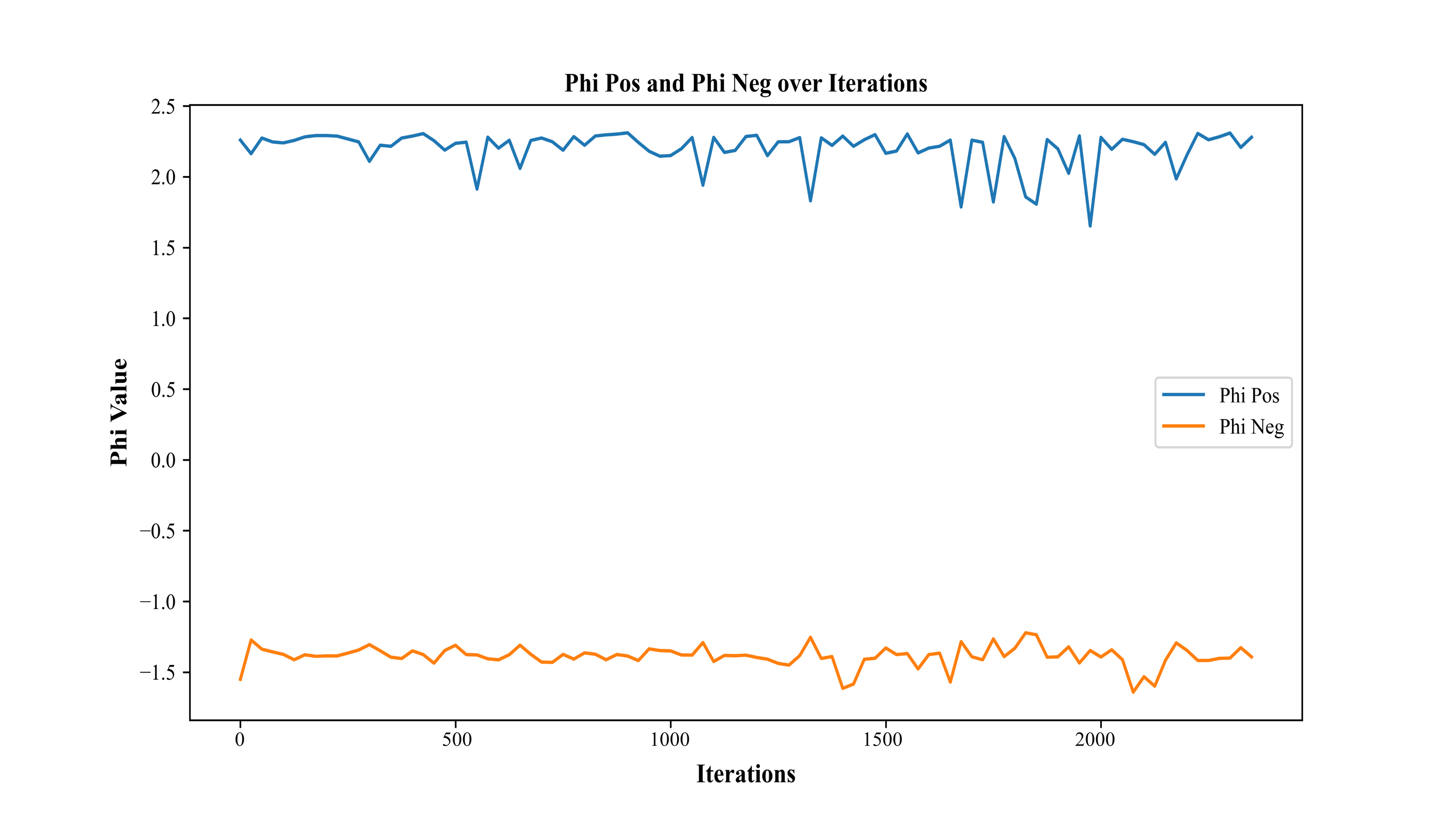}
    \caption{Change in coefficient of LTP and LTD over time on MNIST dataset}
    \label{fig:enter-label}
\end{figure}

 In the context of SNNs, understanding how the subsystems in IIT contribute to consciousness can provide insight into the nature of consciousness in neural networks. The subsystems in IIT include system (The entire system under consideration, which in this case 
 is neural network), partition (The various ways in which the system can be divided into parts or subsystems which in this case are previous and present state of spikes), mechanism (The causal interactions within each partition, describing how elements within each partition interact with each other which in this case is understanding how spikes propagate through the network, how synaptic connections are formed and modified), information (The set of possible states that a system can be in and the probabilities associated with transitioning between those states which in this case is analyzing the information content and how it is integrated across the network). The understanding and working of these subsystems can be visualized through Fig. 4 and Fig. 5. As represented in Fig. 4 the granger causality over iterations, it can be deduced how the neural network's activity (i.e. spike recordings) influenced by the input data. If the spike recordings consistently provide additional predictive power with input data for values of the output spike train. It suggests a causal relationship and provides insights into the cause-effect repertoire of the neural network. Higher Granger causality values from one spike train to another imply a stronger influence. The peaks observed in the graph of the Granger causality values against iterations, signifies moments during training where there are significant changes or fluctuations in the causal relationship. Leading to changes in how it processes information from the input data and optimize its parameters to better capture relevant features in the data. On the other hand Fig. 5 represents the change in values of $\phi_{pos}$ and $\phi_{neg}$ over the iterations. These values represent the positive and negative components of integrated information, respectively and helps in understanding the flow of information within the neural network during learning. $\phi_{pos}$ measures how much information from the input influences the hidden state. A higher $\phi_{pos}$ indicates stronger positive influence and the synaptic weights associated with the connections from the input layer to the hidden layer should be adjusted in a way that strengthens the connections that contribute positively to learning. $\phi_{neg}$ measures how much information from the hidden state influences the input data. A higher $\phi_{neg}$ value implies that adjustments to the synaptic weights associated with the connections from the hidden layer to the output layer should be made to minimize the negative impact on learning. 
By analyzing these values, the network can adapt its synaptic weights to strengthen positive influences while mitigating negative influences, ultimately leading to improved learning and better performance.

\section{Conclusion and Future Work}
 This research work addresses the limitations inherent in existing STDP models by introducing a consciousness driven STDP model. The key factor lies in the information integration and dynamic nature of the proposed $CD$-STDP model, where the coefficients of LTP and LTD are intricately linked to current and past state of the neural activities. To analyse the efficiency of the proposed model on image recognition task, it is implemented with SNN on different benchmark datasets and compared with related work. In addition, analysis of conscious elements and consciousness of the proposed model on the SNN is performed. There are still some extensions in the future for the presented work. First, exploring other ways to define the interplay between consciousness and STDP. Implementing it on more complex and deep SNN structures. Additionally, more biological complexities can be taken into account and the proposed model can be analysed on different applications.

%\begin{thebibliography}{1}
\bibliography{ref}
\bibliographystyle{IEEEtran}

%\end{thebibliography}
\end{document}